%% file: main.tex
\documentclass[conference]{IEEEtran}
\IEEEoverridecommandlockouts
\pdfoutput=1

\usepackage[keeplastbox]{flushend}
\usepackage{hyperref}
\usepackage{cite}
\usepackage{amsmath,amssymb,amsfonts}
\usepackage{algorithmic}
\usepackage{graphicx}
\usepackage{textcomp}
\usepackage{xcolor}
\usepackage{listings}
\def\BibTeX{{\rm B\kern-.05em{\sc i\kern-.025em b}\kern-.08em
    T\kern-.1667em\lower.7ex\hbox{E}\kern-.125emX}}

\usepackage[english]{babel}
\usepackage[utf8]{inputenc}
\usepackage{varioref}
\usepackage{csquotes}
\usepackage[linesnumbered]{algorithm2e}
\usepackage{subcaption}
\usepackage{tikz}
\usepackage[inline]{enumitem}
\usepackage{scalefnt}
\usepackage{siunitx}

\begin{document}

\title{Orchestral: a lightweight framework for parallel simulations of cell-cell communication
\thanks{This work has been funded by the Swedish research council (VR) under
award no. 2015-03964 and by the eSSENCE strategic collaboration of eScience.}
}

\author{\IEEEauthorblockN{1\textsuperscript{st} Adrien Coulier}
\IEEEauthorblockA{\textit{Department of Information Technology} \\
\textit{Uppsala University}\\
Uppsala, Sweden \\
adrien.coulier@it.uu.se}
\and
\IEEEauthorblockN{2\textsuperscript{nd} Andreas Hellander}
\IEEEauthorblockA{\textit{Department of Information Technology} \\
\textit{Uppsala University}\\
Uppsala, Sweden \\
andreas.hellander@it.uu.se}
}

\maketitle

\begin{abstract}

We develop a modeling and simulation framework capable of massively parallel simulation of multicellular systems with spatially resolved stochastic kinetics in individual cells. By the use of operator-splitting we decouple the simulation of reaction-diffusion kinetics
    inside the cells from the simulation of molecular cell-cell interactions occurring on
    the boundaries between cells. This decoupling leverages the inherent scale separation in
    the underlying model to enable highly horizontally scalable parallel simulation,
    suitable for simulation on heterogeneous, distributed computing
    infrastructures such as public and private clouds.
    Thanks to its modular structure, our frameworks makes it possible to couple
    just any existing single-cell simulation software together with any cell
    signaling simulator.
    We exemplify the flexibility and scalability of the framework by using the popular single-cell simulation software eGFRD to construct and simulate a multicellular model of Notch-Delta signaling over OpenStack cloud infrastructure provided by the SNIC Science Cloud.  
\end{abstract}

\begin{IEEEkeywords}
Computational systems biology, high performance and parallel computing, distributed computing, cloud computing 
\end{IEEEkeywords}

\input{Introduction}
\input{Algorithm}
\input{Experiment}
\input{Discussion}
\input{Bibliography}

\end{document}

%% file: Introduction.tex
\section{Introduction}

One important insight gained in computational systems biology is that intrinsic molecular noise in cellular regulatory networks can have important consequences for the functioning of the system \cite{vilar_mechanisms_2002,elf_spontaneous_2004,paulsson_stochastic_2000,sturrock_spatial_2013-2}. This has led to the development of a wide range of methods and simulation software capable of simulating the stochastic dynamics of intracellular kinetics, both in a well-mixed and in a spatial setting \cite{gillespie_perspective:_2013}. 


In the case that one can assume a well-mixed system, i.e. when the rate of chemical reactions can be assumed to be independent of the spatial location of the molecule,  Gillespie's Stochastic Simulation Algorithm is a popular algorithm to generate trajectories of the involved chemical species, who are assumed to follow a continuous time discrete space Markov process. 
As long as the well-mixed assumption holds, this type of modeling has proven very useful, in cell biology but also in other fields where this algorithm
was not primarily intended~\cite{szekely_stochastic_2014}.

Cellular regulation kinetics are inherently spatial though, with reactions frequently taking place in a diffusion-limited regime where the spatial homogeneity assumption breaks down. One popular alternative that simulates stochastic kinetics with spatial resolution is the Reaction Diffusion Master Equation~(RDME)~\cite{elf_spontaneous_2004}.
On an even finer scale, molecules are assumed to be continuously diffusing hard spheres and interact according to Smoluchowski diffusion limited theory. The Green's Function Reaction Dynamics (GFRD) framework~\cite{sokolowski_egfrd_2017} is a popular simulation software in that setting.
There is therefore a well established hierarchy for single-cell modeling
techniques, with ordinary differential equation (ODE) based methods on the
coarser end of the spectrum and detailed particle dynamics methods on the other
end. 

However, in complex multicellular systems, cells do not act in isolation of each other, on the contrary, they interact with each other in many different ways. Through direct contact, where cells physically push or pull each other, changing position and shape over time, and through chemical signaling, where signals are mediated by receptor molecules at the surface of the cells, that trigger a cascade of intracellular reactions when activated, modulating the cell's genetic regulatory networks. Unless simulations are able to capture both these intricate cell-cell interaction mechanisms as well as discrete, spatial stochastic molecular regulation, modelers are missing out on important details.

Several studies have focused on coupling single-cell and multicellular
interactions. Cell mechanics have been modeled with agent-based models~\cite{thorne_combining_2007}, center based models~\cite{meineke_cell_2001},
vertex based models \cite{alt_vertex_2017} or cellular Potts models~\cite{graner_simulation_1992}. These models have
been coupled with ODEs~\cite{delile_cell-based_2017, engblom_scalable_2017}, 
boolean networks~\cite{antoniotti_cognac:_2015}, and more recently the
RDME framework~\cite{engblom_stochastic_2018}. Non-spatial deterministic models
remain the preferred framework due to their relatively low complexity and low computational cost. Similar to the insights gained from using stochastic models in the single-cell setting though,
enabling the use of more detailed stochastic models of chemical kinetics coupled with sophisticated models of cell mechanics interactions could bring significant insight to these models.

Contrary to methods used for single-cell models, for multicellular systems there is no unifying underlying mathematical theory that would
provide an analytical solution to be compared with. Also, there is no well-established model hierarchy, and models are often only
compared qualitatively to each other~\cite{osborne_comparing_2017}. 
So far, cell mechanics models have often been coupled with coarse, simple
internal cell dynamics, such as \emph{ad hoc} rules, ordinary and delayed
differential equations, or simple stochastic dynamics
\cite{gorochowski_agent-based_2016}. It remains, however, unclear how much
information is lost in using these coarse approximations. In other words, using
more detailed models and software for simulating intracellular reaction networks, such as GFRD~\cite{sokolowski_egfrd_2017}, Smoldyn~\cite{andrews_detailed_2010} or URDME~\cite{drawert_urdme:_2012}, could uncover new insights.  Handling such detailed simulation remains highly challenging due to their prohibitive computational cost. 

Given this diversity in both single-cell and multicellular simulation
frameworks, there is a need for tool to easily combine different software into
complex simulations that includes single-cell resolution and cell-cell
signaling. Due to the computational cost, such a framework needs to be able to
leverage modern distributed computing environments such as cloud infrastructure
to seamlessly scale simulations in parallel. This would make quantitative
comparative studies of various model couplings possible, allowing modelers to
determine more easily how much detail is needed in their simulations.

The aim of this paper is therefore to provide such a unifying framework making it possible to construct composite multicellular models from state-of-the-art single-cell simulators and from multicellular modeling frameworks. 
To accommodate as many current and future modeling approaches as possible, the framework needs to be as flexible and general as possible. To achieve this, the only assumption we make regarding our models is that they are possible to split in an operator split fashion. 
Unlike other multiscale agent-based platforms (such as Chaste \cite{mirams_chaste:_2013}
or BSim \cite{gorochowski_bsim:_2012}) our framework is independent of any
language or technology, and focuses on combining different tools into massively
parallel simulations rather than to provide new tools for individual components
of a simulation.
Rather than a new simulation software, or a generic workflow library, our
framework is a tool to compare various combinations of models, systematically
and reproducibly, using existing simulation software
\cite{sokolowski_egfrd_2017}, from the computational biology field on the one
hand, and from parallel computing on the other hand
\cite{dask_development_team_dask:_2016}.



The rest of the paper is organized as follows: first, we introduce our framework
and its components. We then move on to the
scaling study and detail our simple biological test-model and how we have
implemented our framework. We examine both strong and weak scaling of our
algorithm. We conclude by emphasizing the advantages of our approach and
open the way for future developments.

%% file: Algorithm.tex
\usetikzlibrary{calc, arrows, decorations.pathmorphing}

\section{A parallel framework for multiscale simulation of cells}

Our main objective is to design a general framework that makes
it possible to a) combine different existing tools for single cell reaction kinetics simulation and multicellular mechanics models into multi-fidelity multicellular simulations, and b) enable to simulate the resulting model in parallel across a range of platforms. The
simulations should permit the inclusion of spatial details in single cells, and
should be pluggable to allow for the use of  different simulation software for
the different parts of a simulation (multiscale simulation).

Three main design assumptions are critical to meet these requirements. First, we assume that a multicellular simulation can be broken down into principal parts, or simulation modules, e.g: intracellular dynamics, cell signaling, and mechanics.
Second, we assume that
each such simulation module can be simulated one after the other, for a short
timestep~$\Delta t$, such that the numerical solution converges with~$\Delta t
\to 0$. This is true for operator splitting and co-simulation schemes under
reasonably non-restrictive conditions, and a commonly employed strategy in
multiphysics simulation. This assumption enables parallelism, since by
decoupling internal cell dynamics from boundary interactions over the
timestep~$\Delta t$, it allows us to update cells in an embarrassingly parallel
fashion. The amount of parallelism enabled depends on the size of the timestep,
as will be illustrated in the numerical experiments.

Third, we assume that the simulation software used to implement the different simulation modules can be called from the command line, with a single input file and a single output file as arguments (SISO). This assumption makes our framework agnostic to specific technology, software or library to handle the simulation modules. Literally any command line simulator can be used, provided the two above-mentioned assumptions are fulfilled. 

\begin{figure}
    \begin{center}
        \begin{tikzpicture}[<->, thick, >=stealth,
            main node/.style={circle,fill=blue!40, draw,minimum size=0.8cm},
            small node/.style={circle,fill=red!40,draw,minimum size=0.4cm},
            datafile/.style={rectangle,fill=white,draw}]

            \node (center) at (0, 0) {};

            \draw[ultra thick,->] (center) (-10:2cm) arc(-10:-73:2cm);
            \draw[ultra thick,->] (center) (-100:2cm) arc(-100:-168:2cm);
            \draw[ultra thick,->] (center) (-190:2cm) arc(-190:-253:2cm);
            \draw[ultra thick,->] (center) (-260:2cm) arc(-260:-348:2cm);

            \node[main node] (chi) at ([shift=({180:2cm})]center) {$\chi$};
            \node[main node,align=center] (ess) at ([shift=({0:2cm})]center) {$S$};

            \node[small node] (ess2) at ([shift=({90:2cm})]center) {\small{$T_{\chi\to S}$}};
            \node[small node] (chi2) at ([shift=({-90:2cm})]center) {\small{$T_{S\to\chi}$}};

            \node[datafile] (chi.out) at ([shift=({135:2cm})]center) {$\chi$\texttt{.out}};
            \node[datafile] (s.in) at ([shift=({45:2cm})]center) {$S$\texttt{.in}};
            \node[datafile] (s.out) at ([shift=({-45:2cm})]center) {$S$\texttt{.out}};
            \node[datafile] (chi.in) at ([shift=({-135:2cm})]center) {$\chi$\texttt{.in}};

            \node[circle,fill=green!40,draw] (orch) at (center) {O};
            \draw[very thick,->] (center) (-5:1cm) arc(-5:-355:1cm);

        \end{tikzpicture}
        \caption{
            The Orchestral Framework. Blue nodes represent simulation modules,
            while red nodes represent translation scripts.
            Every timestep, the internal cell dynamics are simulated ($\chi$)
            and the result of these simulations is translated to initialize the
            simulation of cell signaling ($S$). This simulation is then run as well
            and the results are translated to update the cell's internal state.
            The central node (O) executes and distributes the execution of every
            program the right order.
        }
        \label{fig:orchestral}
    \end{center}
\end{figure}
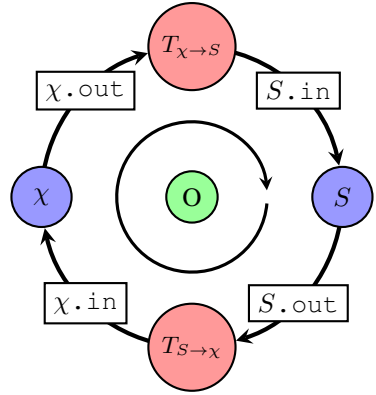

Figure \ref{fig:orchestral} illustrates the architecture of our proposed framework. Blue
nodes represent our two simulation modules, one for internal spatial
details~($\chi$), and the other for external cell interactions~($S$). Red nodes
are the programs responsible for providing cell interactions with summarized
internal spatial data and for updating internal dynamics with information from
cell-cell interactions, essentially coupling the simulation modules together.
The green node in the center is in charge of executing these modules in the
right order and as efficiently as possible.
For every timesteps, each module is executed once, in the order following the
thick arrow. The simulation ends once all the timesteps have been executed.

We point out that all the above requirements are very weak assumptions that are
easily satisfied. On the one hand, the splitting should work provided the
coupling implemented by the user (i.e.~the way data are summarized and shared
between simulation modules) is correct and makes senses, physically speaking.
On the other hand, even if simulation software do not always provide a SISO
interface, it is very simple and straightforward to wrap them into a small
script that would ensure compatibility.

In the following subsections we describe the main components of the framework.
Our implementation is available at \url{https://bitbucket.org/Aratz/orchestral/}.

\subsection{Simulation modules}
In this
paper, we restrict the scope to study simulation modules for intracellular dynamics and cell-cell signaling.


\subsubsection{Internal cell dynamics module}
This module focuses on simulating chemical kinetics inside cells, e.g. the
interaction between proteins and genes. In this paper it will be denoted by the
Greek letter $\chi$.
There is a wide range of models that can be used here, from ordinary differential
equation models to more detailed particle based algorithms.
One important remark here is that the role of the internal dynamics module is to update the internal state of independent cells. All types of molecular interactions involving more than one cell will be handled
in the next module. This makes the update of cell internal dynamics across a cell population embarrassingly parallell over the splitting timestep $\Delta t$..
\subsubsection{Cell signaling module}
Here pairwise signaling is considered, that is, all molecular interactions
involving exactly two cells. This can for example be binding of a ligand in one
cell to a receptor at an adjacent cell. All pairs being independent, this step
is also naturally embarrassingly parallel. We will use the letter $S$ to
represent this module.

\subsection{Translation scripts}
The translation scripts connect the simulation modules. Before a simulation
module is run, an input file is generated by the translation script from the
data computed previously by the other simulation modules
(Figure~\ref{fig:orchestral}). The translation scripts ensure compatibility between the
simulation modules. If such a module is replaced by a new one using a different
data format, the only action needed in this case is then to update the
translation scripts.

Specifically, when translating from an internal dynamics model to a signaling model
($T_{\chi\to S}$), the script takes a pair of cells as input. The script
can then be used to filter out molecules that are close to the common
boundary of the two cells and put them in the signaling model input file. When
translating from a signaling model to an internal dynamics model
($T_{S\to\chi}$), the script takes an unspecified number of signaling pairs as
parameters, which depends on the number of neighbors of the target cell.
Typically, this script is used to propagate new molecules from the signaling
model back into the internal dynamics model.

\subsection{Orchestration}
\lstset{
    string=[s]{"}{"},
    stringstyle=\bfseries,
    comment=[l]{:},
    breaklines=true,
    breakatwhitespace=true,
    postbreak=\mbox{$\hookrightarrow$\space},
}

\begin{figure}
\begin{lstlisting}
{
 "0": {
  "position": [0.0, 0.0, 0.0]
  "neighbors": [1, 2],
  "epsilon": 0.1,
  "wsize": 1e-06,
  "seed": 1,
  "end_time": 0.1
 },
 "1": {
  "position": [1e-06, 0.0, 0.0]
  "neighbors": [0, 4],
  "epsilon": 0.1,
  "wsize": 1e-06,
  "seed": 2,
  "end_time": 0.1
 },
 "2": {
  "position": [0.0, 1e-06, 0.0]
  "neighbors": [4, 0],
  "epsilon": 0.1,
  "wsize": 1e-06,
  "seed": 3,
  "end_time": 0.1
 },
 "4": {
  "position": [1e-06, 1e-06, 0.0]
  "neighbors": [2, 1],
  "epsilon": 0.1,
  "wsize": 1e-06,
  "seed": 5,
  "end_time": 0.1
 }
}
\end{lstlisting}
    \caption{
        Network file. This file describes the position as well as specific
        parameters and neighbors of each cells.
    }
    \label{fig:network}
\end{figure}

\begin{figure*}
\begin{lstlisting}
{
    "cell_executable": "modern_egfrd/build/bin/RunGfrd --custom -seed {seed} -e {end_time} -wsize {wsize} -in {input_file} -out {output_file}",
    "signaling_executable": "python3 signaling/signaling.py -seed {seed} -e {end_time} -in {input_file} -out {output_file}",
    "translation_X2S":"python3 translation_X2S.py {network_file} {cell_output_files} {signaling_input_file}",
    "translation_S2X":"python3 translation_S2X.py {network_file} {target_cell_output_file} {signaling_files} {target_cell_input_file}",
    "data_folder":"data/",
    "n_steps":100
}
\end{lstlisting}
    \caption{
        Configuration file. This file contains all the information necessary to
        run the four modules, i.e. command line syntax and parameters, as well
        as global simulation parameters.
    }
    \label{fig:config}
\end{figure*}
Finally, the main module of our framework is the orchestrator, which is responsible for executing simulation modules and translation scripts, and for
efficiently distributing the computations across the computing grid. Again, this module is completely independent from the simulation modules. Replacing a simulation
module does not affect the orchestrator, and replacing the orchestrator (for
instance to fit the computing cluster of the user's choice better) does not
affect the implementation of the simulation modules.

\begin{figure}
    \begin{center}
        \includegraphics[width=\columnwidth]{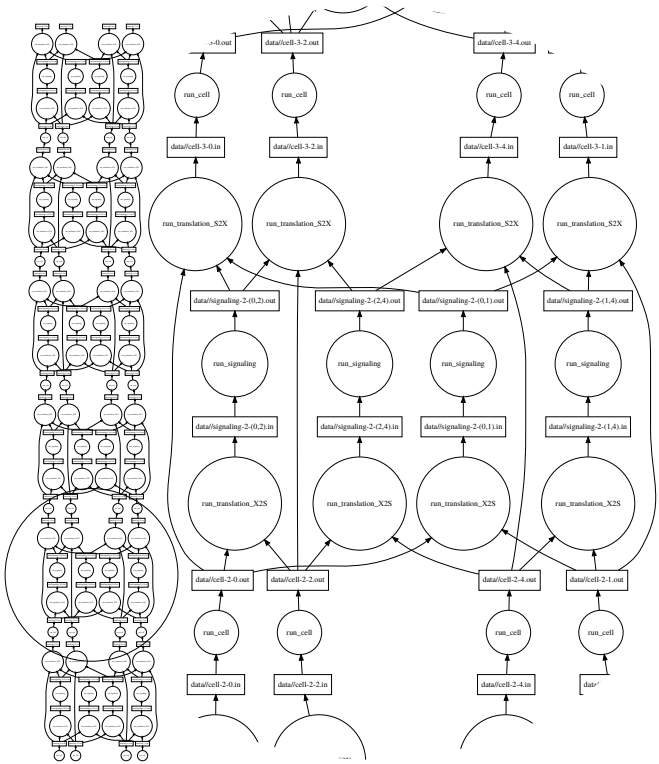}
        \caption{Task graph generated by Dask, for a model involving four cells
        and six timesteps. Round nodes are computation and translation steps,
        while rectangular nodes are input and output files.}
        \label{fig:dask}
    \end{center}
\end{figure}

We chose to use Dask~\cite{dask_development_team_dask:_2016},
a python library for parallel and distributed computing. However, we point out
that this is only one of many available alternatives, and that switching
to any other library could be done very easily. In Dask, tasks can be described
and hierarchized as a directed acyclic graph (DAG) using a simple Python
dictionary, which makes the code very simple to understand. Switching between
local multicore simulation and distributed computations on a high performance
cluster or a cloud is done by changing a single line in the orchestrator and updating
the configuration file so that the data are written in the appropriate shared
area. Figure \ref{fig:dask} shows an example of such a graph for one of our
simulations, where circles are tasks and rectangles are input and output files.

The orchestrator takes two input files as parameters. The first file describes
the cell network, i.e. the position of each cell, the id of their neighbors,
and model specific simulation parameters. An example of such a file, in
\texttt{json} format, is given in Figure~\ref{fig:network}. The data is
encoded as a dictionary, where unique cell identification
numbers are associated to their parameters. It is possible to
fine tune the parameters so as to have various models for each cell. Only
\texttt{position} and \texttt{neighbors} are used by the orchestrator, all
other parameters are fed to the internal dynamics module when the simulation is
launched.

The second file is a configuration file that provides the syntax of the
command line programs used to execute the simulation modules and translation script, as well as global simulation parameters, such as the number of time steps to run. An example
of such a file is given in Figure~\ref{fig:config}. Each command line is written so that it is easy to format it using Python's \texttt{format} method by
simply feeding the simulation parameters to this function. Changes to the format of the configuration file only mandates updates of the orchestrator module, but not the simulation modules or translations scripts. 



%% file: Experiment.tex
\section{Results}
\label{s:results}
In order to demonstrate the usefulness of our framework and to test its scalability, 
we implement a simplistic but biologically realistic model of cell-cell signalling based on the single-cell simulation software eGFRD \cite{sokolowski_egfrd_2017}. The objective here is to provide a proof-of-concept model useful to study parallel scalability and to demonstrate how to use the framework. We conduct strong scaling experiments on a modern multicore machine and weak scaling experiments over a virtual cluster deployed in an OpenStack community cloud. 

\subsection{Setup}

\subsubsection{Model}
\begin{figure}
    \begin{center}
        \includegraphics[width=\columnwidth]{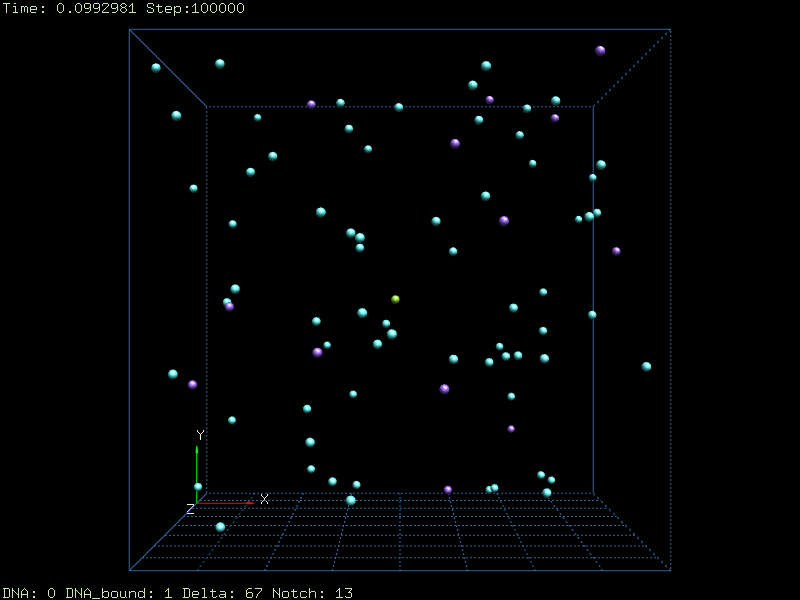}
        \caption{eGFRD single-cell simulation screenshot \cite{sokolowski_egfrd_2017}.
        Cyan spheres represent Delta molecules while purple spheres represent Notch molecules.
        The green sphere in the center is a bound DNA promoter site.}
        \label{fig:egfrd}
    \end{center}
\end{figure}

The model implemented is a simplified model of the Notch signaling pathway \cite{shaya_notch_2011}. The cells' membranes
contain $Notch$ receptors and $Delta$ ligands (molecules binding to the
receptor). When cells are in contact, $Notch$ and $Delta$ proteins can bind and
this will release $Notch$'s intracellular domain ($NICD$) into the cell. $NICD$
will then diffuse into this cell and potentially influence gene expression. In
our model, $NICD$ acts as a repressor of $Delta$, i.e. it suppresses the
creation of new $Delta$ protein. A single DNA molecule is placed in the center
of each cell, modeling the $Delta$ promoter. When not occupied by $NICD$, it
transcribes $Delta$ at a rate $t_D$. $Delta$ proteins then diffuse in the
cells. Figure \vref{fig:egfrd} shows a snapshot of this system when simulated
with eGFRD, where each molecule is approximated by a hard sphere diffusing in a
cube representing the cell. DNA is shown in green in the center of the cube,
while $Delta$ molecules are cyan and $NICD$ molecules are purple. $NICD$ molecules modulate $Delta$ transcription upon binding to DNA.

\begin{table}
    \begin{center}
        \begin{tabular}{lr}
            Parameter & Value\\
            \hline
            Diffusion Constant, $D$ & \SI{1e-12}{\square\meter\per\second}\\
            Transcription, $t_D$ & \SI{1e2}{\per\second}\\
            DNA binding, $k_a$ & \SI{1e-18}{\cubic\meter\per\second}\\
            DNA unbinding, $k_d$ & \SI{2.5e1}{\per\second}\\
            Delta to $NICD$, $k_N$ & \SI{1}{\per\second}\\

        \end{tabular}
        \caption{
            Model parameters used for our simulations. Since the focus of this
            paper is on performance rather than on biological realism, it is
            important to stress that these parameters where not choosen to be
            realistic, but rather so that the model would exhibit some
            interesting behavior to demonstrate the utility of our approach.
        }
        \label{tab:parameters}
    \end{center}
\end{table}

To model cell-cell signalling, we assume that when $Delta$ molecules are within a distance $\epsilon$ of the cell membrane (boundary of the cube), they can be mirrored to the cell on the other side of the membrane
and turn into NICD at rate $k_N$. NICD then diffuses inside the receiving cell and
potentially associates to DNA at rate $k_a$. This binding turns off $Delta$
transcription. NICD unbinds from DNA in bound state at rate $k_d$. Except DNA,
all species degrade at the same rate $d$. The parameter values used in our
simulation are summed up in Table \ref{tab:parameters}. These values are
We tuned the model to illustrate our framework with interesting features. In
the case one would want to perform a more detailed modeling study, one would
need to take values from the literature.

We model cells as identical cubes, arranged in a 2D layer. Hence, each cell has at most four neighbors.

The chemical reaction turning $Delta$ into $NICD$ is handled by the signaling
model, while all other reactions are handled as internal cell dynamics.
Coupling is done by transferring all $Delta$ molecules within $\epsilon$ of a
\emph{single} face of the cube to the associated signaling model. Coupling in
the other direction is done by simply reporting the molecules from the
signaling model back into the appropriate cell.

\begin{figure}
    \begin{center}
        \includegraphics[width=\columnwidth]{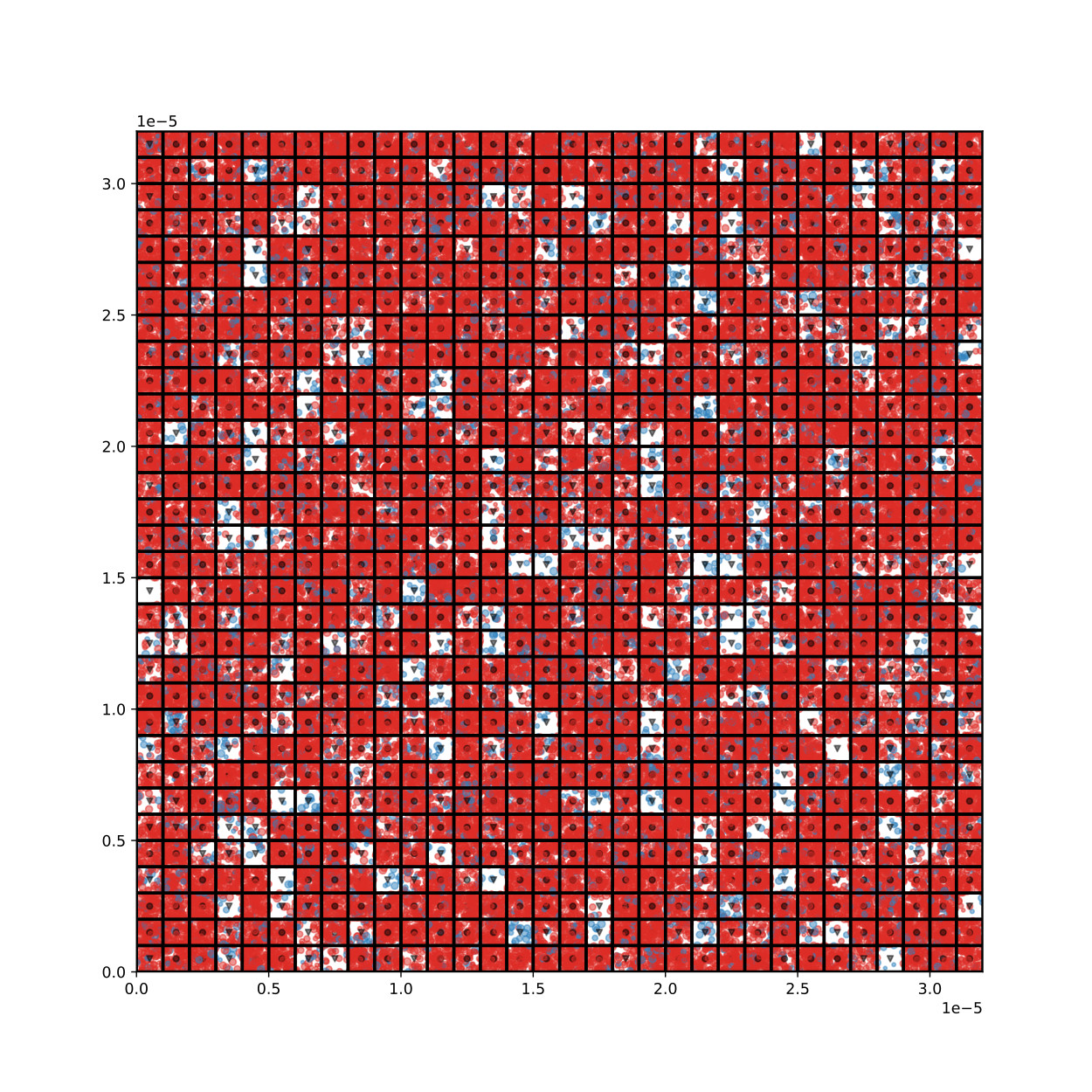}
        \caption{Snapshot of a patch of 1024 cells simulated by our algorithm after 10
        timesteps. Red circles represent Delta molecules, while blue circles
        represent NICD molecules. DNA is represented by black circles (free
        DNA) and black triangles (suppressed DNA). As can be seen, DNA suppression causes
        the cell to stop producing $\Delta$ and to become empty.
        }
        \label{fig:cell_patch}
    \end{center}
\end{figure}

Figure \vref{fig:cell_patch} shows the output of a simulation of 1024 cells
after 10 timesteps, each red dot is a $Delta$ molecule, each blue dot is a
$NICD$ molecule, and DNA is shown in black, either in free state (circles)
or bound state (triangles). Suppressed cells show a loss in $Delta$ production
and appear as white, while cells with a free DNA promoter continue to produce $Delta$
and remain red. Under some specific parameter regimes, this model is known to
exhibit a so-called checkerboard pattern, where free cells are arranged in
a regular pattern. The more noisy the system becomes, the less regular the
pattern~\cite{sprinzak_cis-interactions_2010}.

\subsubsection{Implementation}
We implement the single-cell model using eGFRD \cite{sokolowski_egfrd_2017}. We
use its custom simulation feature to both describe our model and set up reading
and writing input and output files. The signaling model is implemented with a
custom made python script. All translation scripts are implemented as python
scripts.

\subsection{Scaling}

A key feature of the framework, as implemented using Dask, is the ability
to seamlessly leverage a wide range of parallel and distributed computing resources.
To demonstrate this flexibility we here tested the performance and scalability when executed on a single high-end multicore machine and b) using variable-size virtual distributed Dask-clusters deployed over cloud infrastructure as a service (IaaS) in the SNIC Science Cloud (SSC) \cite{toor_snic_2017}.
For the multicore machine, a single node of the Rackham system provided by the Uppsala
Multidisciplinary Center for Advanced Computational Science (UPPMAX) was used. The used node had two 10-core Xeon E5-2630 V4 processors running at 2.2 GHz
(turbo 3.1 GHz). The SSC is a community cloud based on OpenStack, and in the same manner as public cloud providers such as AWS and Azure, it provides
general infrastructure as a service (IaaS) to the Swedish research community. 

We adapted the MOLNs
orchestration toolkit \cite{drawert_molns:_2016} from the StochSS \cite{Drawert2016StochasticBiologist} suite to deploy virtual Dask clusters of
variable sizes (\url{https://github.com/ahellander/molns/orchestral-dask}). MOLNs deploy an SSHFS shared filesystem which we use to share all the
intermediate input-output data files of the simulation.

\begin{figure}
    \begin{center}
        \includegraphics[width=\columnwidth]{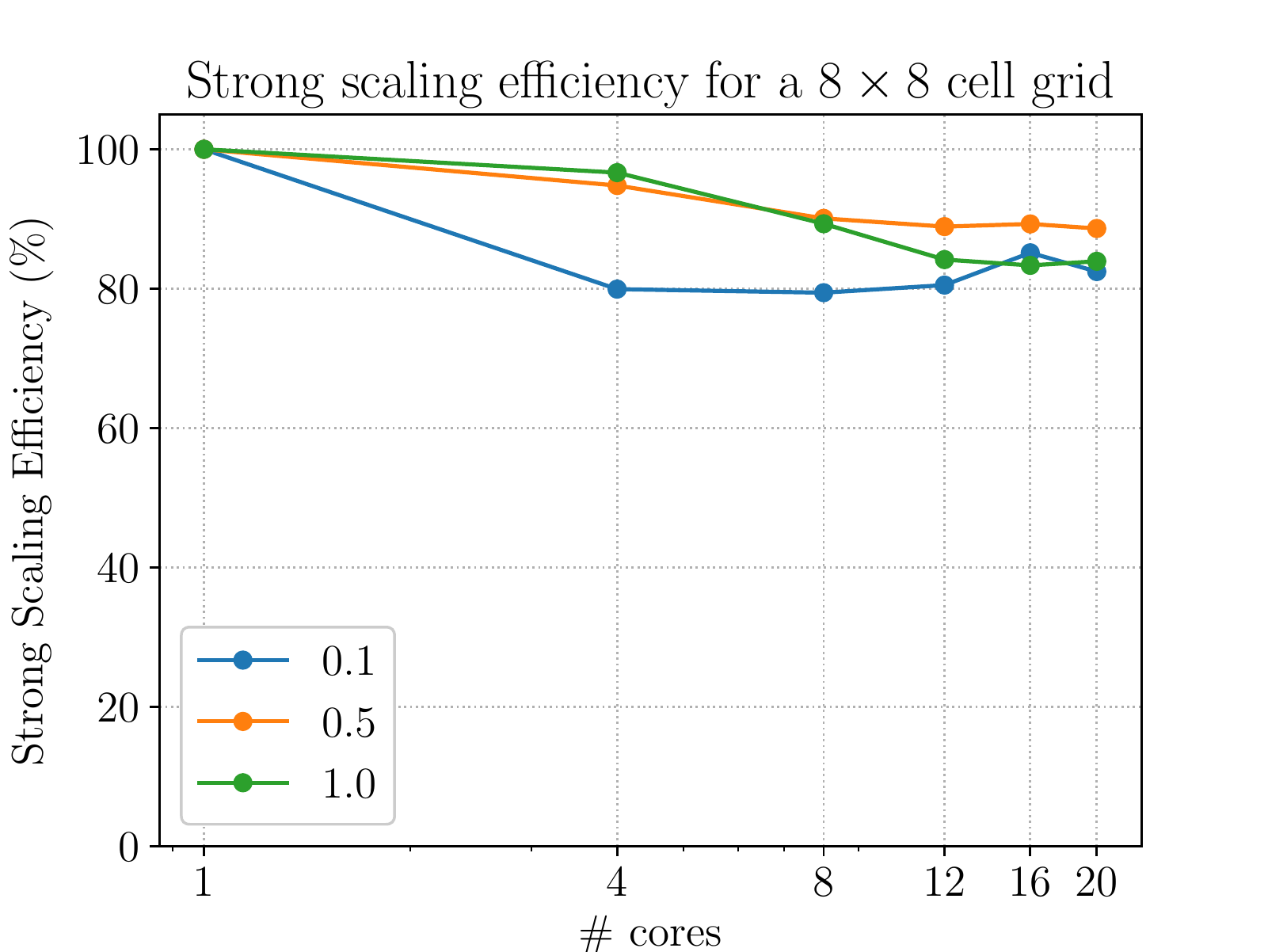}
        \caption{
            Single node strong scaling study of a 64-cell grid. The simulation
            is run for 10 seconds, with timestep length set to 1.0, 0.5 and 0.1
            second. Our framework scales smoothly up to all 20 cores of our
            machine, maintaining about 80\% strong scaling efficiency for
            all setups.
        }
        \label{fig:strong_scaling}
    \end{center}
\end{figure}

We begin with a strong scaling study on a single node, where we study the
computation time as a function of increased number of cores with a fixed model size. We simulate an $8\times 8$ grid of cells for ten seconds simulation time. We set up Dask to
use the threaded single machine scheduler and ran the simulation on 1, 4, 8,
12, 16 and 20 cores. Figure~\vref{fig:strong_scaling} shows the parallel efficiency for splitting timesteps $\Delta t$ of $0.1$, $0.5$ and $1.0$ second, respectively.

Using a single core, the simulation took between 40 and 140 minutes. Using the
full 20 cores of the shared memory node, the simulation took between 2 and 8
minutes, depending on $\Delta t$, thus demonstrating the usability of our
framework
for modeling purposes. Our framework hence shows very good parallel efficiency
for all timesteps, with at slightly more than $80\%$ parallel efficiency, with maximum efficiency obtained with $\Delta t = 0.5$. This can be explained by the fact that with $\Delta t = 1.0$, the task granularity is coarse due to the relatively low number and long duration of the tasks.
With $\Delta t=0.1$, on the other hand, the tasks become too short and
the scheduling overhead of Dask leads to a loss of performance.

\begin{figure}
    \begin{center}
        \includegraphics[width=\columnwidth]{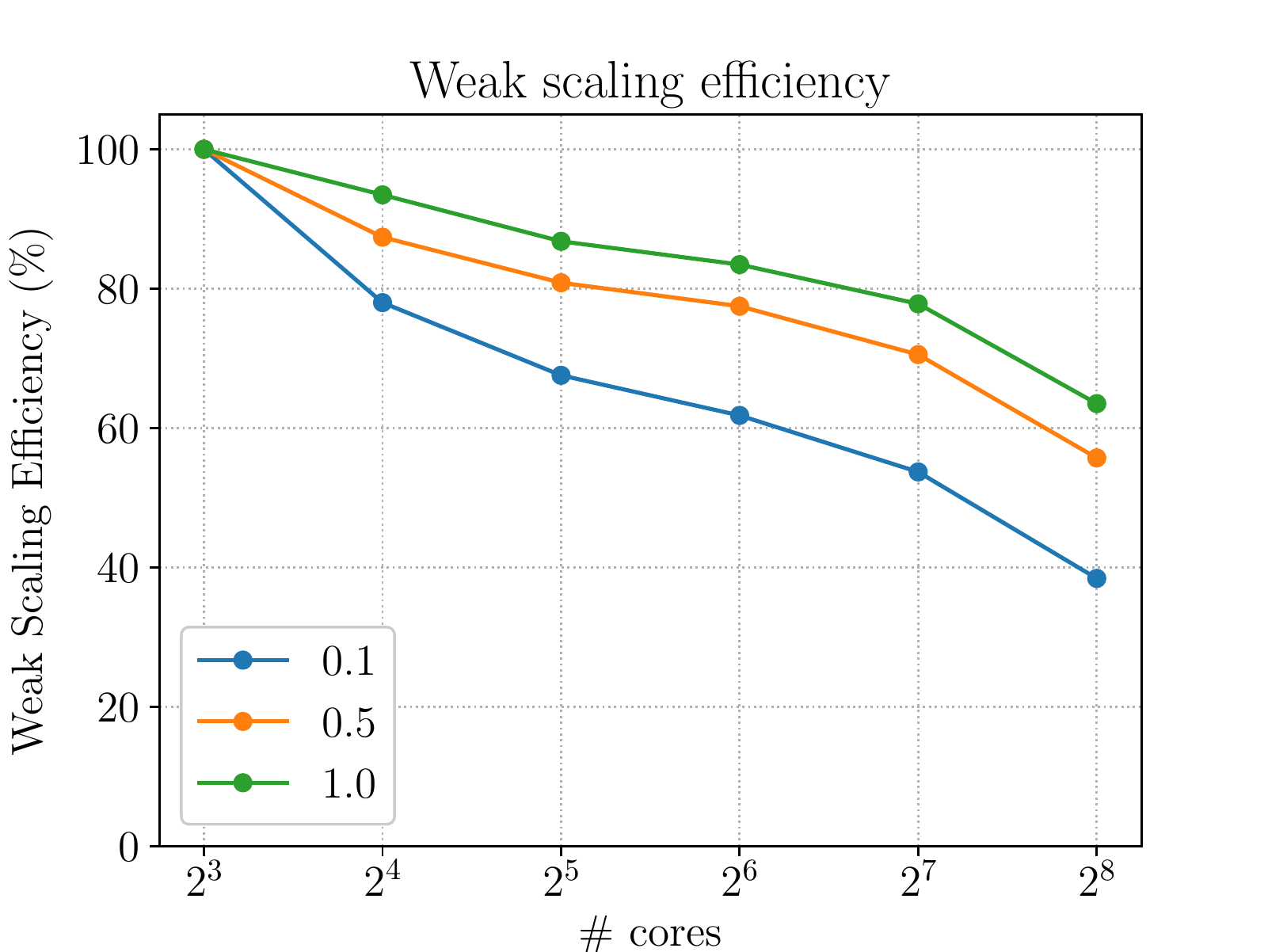}
        \caption{Weak scaling study. Each simulation consists of 100 cells
        per node and are run for 10 timesteps. We study three cases with
        timestep length ranging from 0.1 to 1.0. The results confirm the
        performance of our framework up to the limits of our cluster.}
        \label{fig:weak_scaling}
    \end{center}
\end{figure}

Next we assessed the weak scaling properties of our implementation. We used $ssc.xlarge$ VM instances each with 8 VCPUs and 16GB RAM and kept the number of cells simulated per core in the cluster constant as we scaled the size of the distributed cluster. We start
from a grid of 100 cells for one VM (8 VCPUs), up to 3200 cells on 32 VMs
(256 VCPUs). We run the simulation for 10 timesteps.

\begin{figure}
    \begin{center}
   \includegraphics[width=0.7\columnwidth]{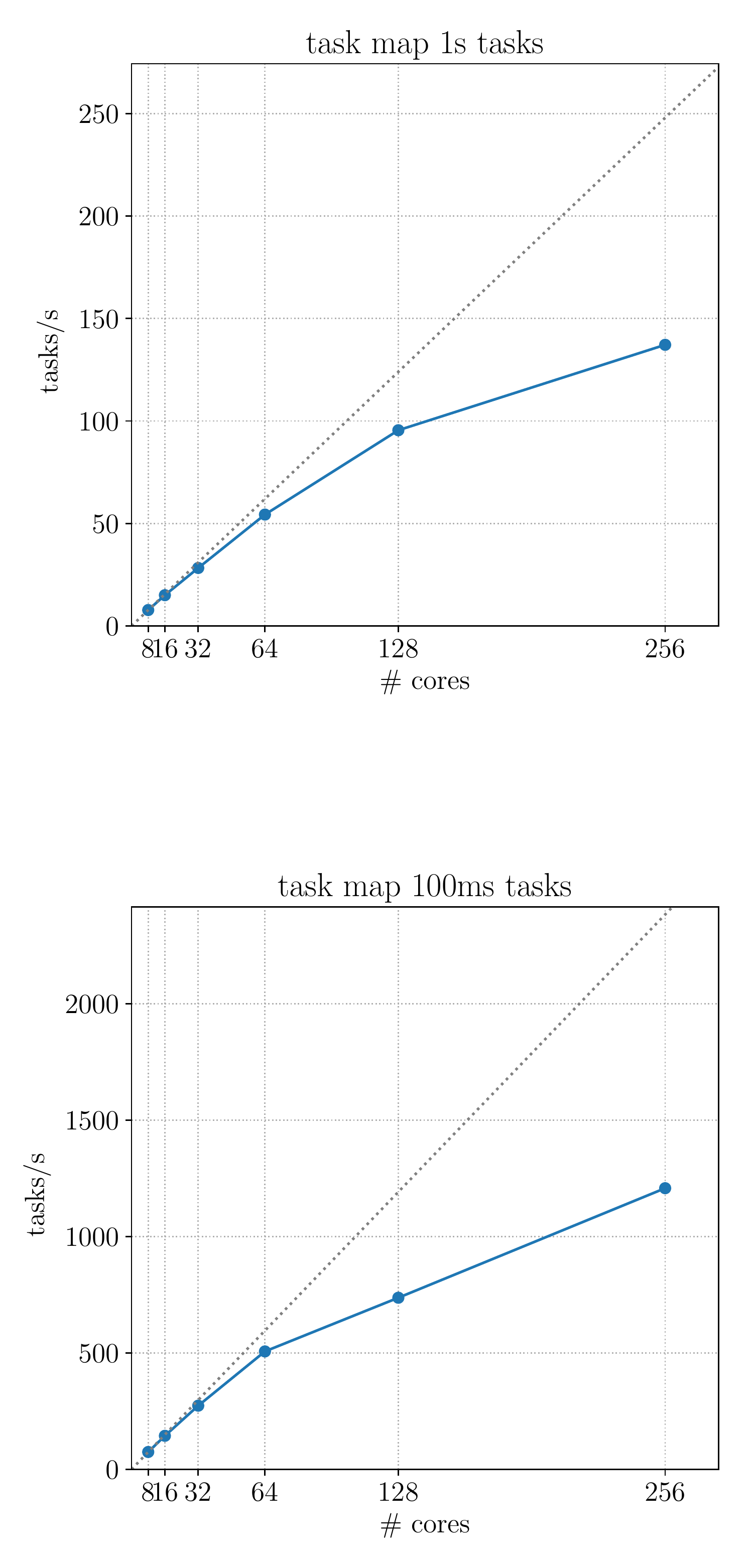}
        \caption{Weak scaling baseline benchmark focusing on Dask performance for deterministic map tasks of 1s processing time (top) and 100ms processing time (bottom).}
        \label{fig:latency}
    \end{center}
\end{figure}

Again, our experiments demonstrate good parallel efficiency, with more
than 50\% efficiency across all tested timesteps on 16 VMs. Using 32 VMs, the efficiency drops as we approach the hardware limits of the cluster underlying the private cloud. It should be noted that the weak scaling properties depends on $\Delta t$, with the efficiency decreases with decreasing timestep. This is explained by the fact that as the timestep is reduced, the simulation tasks become shorter and shorter and  scheduling and communication overheads 
limit scalability for the distributed system. In fact, to establish a baseline performance of Dask as deployed on the SSC virtual infrastructure we ran a publicly available Dask weak scaling benchmark\footnote{http://matthewrocklin.com/blog/work/2017/07/03/scaling} measuring the throughput for simple deterministic tasks. 
In this benchmark, each task consists of simply incrementing an integer, and waiting
for a small delay (\SI{1}{\second}, \SI{100}{\milli\second}). The
gray line shows the ideal performance. As can be seen, Dask over SSC VMs achieve close to ideal scaling up to 64 VCPUs, or 8 VMs, for both task granularities, with 74\% and 54\% efficiency for 128 and 256 cores respectively for \SI{1}{\second} tasks. For \SI{100}{\milli\second} task, the corresponding figures are  57\% and 47\% respectively. These results are consistent with results in the published benchmark with clusters deployed over the Google container platform. 

The weak scaling efficiency of our multicellular simulations with Orchestral shown in Figure \ref{fig:weak_scaling} should be seen in this context. For $\Delta t = 0.1$, individual tasks associated with updating the internal state of cells by running eGFRD vary due to stochasticity and particle counts in individual cells, but upon manual inspection in the Dask status UI they are found to vary in the interval \SI{500}{\milli\second} -- \SI{2}{\second}, with the signaling module tasks taking around 300 -- \SI{600}{\milli\second}/task. This means that Orchestral achieves a weak scaling efficiency close to the ideal case (for this orchestrator and cloud infrastructure combination), demonstrating the lightweight nature of our framework. 

%% file: Discussion.tex
\section{Discussion and conclusion}
In this paper, we have presented a new framework for constructing and simulating high-fidelity
models of multicellular systems from existing frameworks for single-cell simulation. The driving motivation for Orchestral is a
need to be able to conveniently combine the many existing frameworks for
single-cell resolution reaction-diffusion models with the diverse landscape of
models of cell mechanics. When combining these modeling levels, they need to be
coupled via molecular cell-cell signaling. Orchestral provides a model for
implementing such coupling, and for simulating the resulting model massively in
parallel over a wide range of distributed computing environments. This is
important, since the models resulting from integrating single-cell simulation
with multicellular simulation become very computationally expensive. In this
paper, we demonstrated the potential of Orchestral by turning the popular
single-cell simulation software eGFRD~\cite{sokolowski_egfrd_2017} into a
multicellular model where cells signal each other via a simplistic model of the
Notch-Delta pathway over adjacent boundaries of individual cells. 

The parallel scalability was demonstrated both in a strong scaling experiment on a shared memory system, and in weak scaling
experiments over a multi-tenant science cloud IaaS infrastructure. The experiments serve to highlight the dependency of parallel efficiency on the splitting time step $\Delta t$, since this timestep determine the task sizes and hence latency requirements of the combination of underlying orchestrator and infrastructure. Since the choice of timestep depends on the needed and attained accuracy of the simulation (which is model dependent) it is not easy to \emph{a priori} determine what type of computing platform is most suitable to accelerate simulation. In future work, we plan to study this accuracy/performance trade-off for specific combinations of models and single-cell simulation software, in order to equip orchestral with highly performant orchestrator implementations for a range of timestep requirements and computing infrastructure.  

Orchestral is primarily aimed at modelers with basic scripting experience. Due to
its modular design, it is easy to reuse existing parts and patterns from a
previous successful set up. 
Our framework puts no constraint on the technology used by the
solvers, the translation scripts, or the orchestrator.  All these modules are
independent programs and can be exchanged at will.  Indeed, replacing one
simulation module only introduces the requirement to adapt the translation scripts to the file format of the new solver. In general, this is usually easily
done as it mostly entails filtering and reformating the data. 


Finally, the orchestrator engine is also easily replaceable for a user with
experience in distributed parallel computing. It is our intention for
Orchestral to evolve with the help of the community to support a wide range of
orchestrators tuned to specific platforms, for example optimized for shared
memory desktop computers, for traditional HPC batch computing schedulers, and
for virtual clusters deployed in private and public clouds. Writing a new
orchestrator does not require any specific knowledge about the actual
simulation modules or translation scripts. Therefore, programmers can focus on
performance, without worrying about model compatibility. We found Dask to
provide a good balance between simplicity, portability and performance, and
chose to use it to implement the orchestrator that ships with Orchestral.
There are however many alternatives available~\cite{augonnet_starpu-mpi:_2012},
\cite{tejedor_clusterss:_2011}, \cite{bosilca_parsec:_2013},
\cite{zaharia_apache_2016}, \cite{rubensson_chunks_2014}, \cite{zafari_ductteip_2018},
\cite{lampa_towards_2016},
which could potentially help us to push performance for low-latency
requirements on distributed infrastructures. In fact, Orchestral could be used
to create a comparative benchmark of all these libraries.

Here we have presented a proof of concept of our framework based on eGFRD as single-cell simulator. Future work on the modeling side  involves exploring and comparing new combinations of simulators, in particular adding a module for cell-mechanics. 


Finally, within our setup it is already possible to use different models to update 
different areas of the cell population, e.g. to use highly detailed simulation modules (such as eGFRD) at
the boundary of a tissue (or generally where cells are actively engaged in signaling) and
coarser modules (such as the RDME or even well-mixed models) in areas with less activity. In the future we will explore adaptive multiscale methods to switch between such models automatically. 

\section*{Acknowledgments}
The authors would like to thank S. Mathias and F. Coulier for constructive criticism of the manuscript.
This work has been funded by the Swedish research council~(VR) under award
no.~2015-03964 and by the eSSENCE strategic collaboration of eScience. Cloud
computing resources were provided by the Swedish National Infrastructure for
Computing via the SNIC Science Cloud.  Other computations were performed on
resources provided by SNIC through Uppsala Multidisciplinary Center for
Advanced Computational Science~(UPPMAX) under project~SNIC~2018/8-157.

%% file: Bibliography.tex
\bibliographystyle{plain}

\bibliography{Bibliography}